\documentclass[aps,prb,twocolumn,showpacs,floatfix,superscriptaddress]{revtex4-1}
\usepackage{graphicx,epsfig,color,dcolumn}
\usepackage{ulem}

\begin{document}

\title{The bandstructure of gold from many-body perturbation theory}

\author{T. Rangel}
\email[Email: ]{tonatiuh.rangel@cea.fr\\ Present address: CEA, DAM, DIF, F-91297 Arpajon, France}
\affiliation{European Theoretical Spectroscopy Facility (ETSF)}
\affiliation{Institute of Condensed Matter and Nanosciences (IMCN), Universit\'e Catholique de Louvain, Chemin des \'Etoiles 8 bte L7.03.01, B-1348 Louvain-la-Neuve Belgium}

\author{D. Kecik}
\affiliation{Paul Scherrer Institut, Materials Science and Simulation,
NUM/ASQ, CH-5232, Villigen, Switzerland}
\affiliation{Ecole Polytechnique F\'ed\'erale de Lausanne (EPFL),
Institute of Materials (IMX) CH-1015, Lausanne, Switzerland}

\author{P. E. Trevisanutto}
\affiliation{European Theoretical Spectroscopy Facility (ETSF)}
\affiliation{National Nanotechnology Laboratory (NNL), Istituto Nanoscienze-CNR, Lecce, Italy}
\affiliation{Institut N\'eel, CNRS and UJF, Grenoble, France}

\author{G.-M. Rignanese}
\affiliation{European Theoretical Spectroscopy Facility (ETSF)}
\affiliation{Institute of Condensed Matter and Nanosciences (IMCN), Universit\'e Catholique de Louvain, Chemin des \'Etoiles 8 bte L7.03.01, B-1348 Louvain-la-Neuve Belgium}

\author{H. Van Swygenhoven}
\affiliation{Paul Scherrer Institut, Materials Science and Simulation,
NUM/ASQ, CH-5232, Villigen, Switzerland}
\affiliation{Ecole Polytechnique F\'ed\'erale de Lausanne (EPFL),
Institute of Materials (IMX) CH-1015, Lausanne, Switzerland}

\author{V. Olevano}
\affiliation{European Theoretical Spectroscopy Facility (ETSF)}
\affiliation{Institut N\'eel, CNRS and UJF, Grenoble, France}

\date{\today}

\begin{abstract}
The bandstructure of gold is calculated using \textit{ab initio} many-body perturbation theory.
Different approximations within the $GW$ approach are considered.
Standard single shot $G_0W_0$ corrections modify the $sp$-like bands while leaving unchanged the 5$d$ occupied bands.
Beyond $G_0W_0$, quasiparticle self-consistency on the wavefunctions lowers the 5$d$ bands.
Globally, many-body effects achieve an opening of the 5$d$-6$sp$ interband gap of $\sim$0.4 to $\sim$0.8~eV, reducing the discrepancy with the experiment.
Finally, the quasiparticle bandstructure is compared to the one obtained by the widely used HSE (Heyd, Scuseria, and Ernzerhof) hybrid functional.
\end{abstract}

\maketitle

\section*{Introduction}
\begin{figure}[h!]
\includegraphics{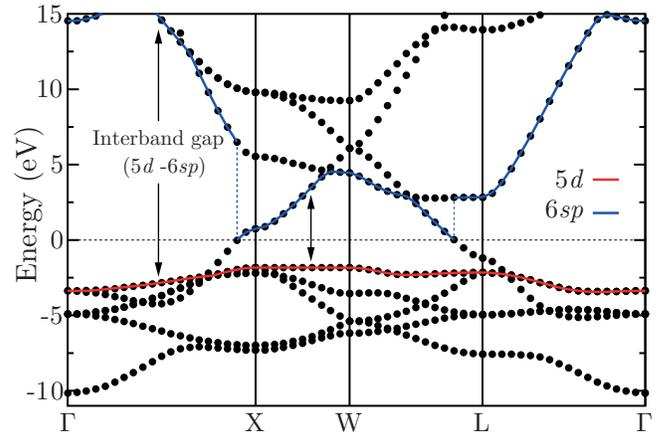}
\caption{(Color online)
Gold DFT-PBE scalar-relativistic bandstructure (black points).
The Fermi level is set to 0 (dashed-black line).
Red (grey) lines disentangle 5$d$-character topmost occupied bands, while blue (black) lines indicate 6$sp$-like lowmost empty bands.
The arrows show the interband gap between the highest occupied $5d$ band and the lowest unoccupied $6sp$ bands.}
\label{fig:interband}
\end{figure}

The theoretical determination of the bandstructure of gold has been an open issue for more than four decades.
Early works from the 70s~\cite{christensen.1971,heimann.1977,pyykko.1979} focused on relativistic effects which are responsible for its yellow color.
Thereafter, the band structure calculated by Christensen and Seraphin~\cite{christensen.1971} has been used as a reference to interpret photoemission experiments.
More recently, a few discussions on this topic appeared in the literature.
The cohesive energy in noble metals was shown to contain large terms arising from dispersion forces, such as van der Waals interactions,~\cite{maggs.1987} pointing to the importance of many-body correlations for closed shell $d$ electrons.
Newer experimental~\cite{courths.1986} and theoretical~\cite{romaniello.2005} works confirmed previous findings.~\cite{heimann.1977}
The gold bandstructure, calculated by density functional theory~(DFT) within the local density approximation~(LDA) or the generalized gradient approximation~(GGA),
presents an underestimation of the 5$d$-6$sp$ interband gap (see Fig.~\ref{fig:interband}) by $\sim$ 1.0~eV with respect to the available experimental data.
Similar discrepancies were encountered for other noble metals.
To solve these disagreements, quasiparticle (QP) corrections to the DFT eigenvalues have been applied with great success.
For instance, in silver and copper, the non-self-consistent $G_0W_0$ approach corrects the DFT interband gap in a remarkably good agreement with the experiments.~\cite{marini.2001,marini.2002,rohlfing.2010}

In fact, the standard $G_0W_0$ approach ({\it i.e.}, starting from DFT) relies on the assumption that the QP wavefunctions are close to the DFT ones.
In some cases, such as for the $3d$ electrons in vanadium dioxide,~\cite{gatti.2007} this hypothesis does not hold.
Two schemes have thus been proposed in order to go beyond standard $G_0W_0$ by introducing an update of the wavefunctions towards self-consistency:
on the one side, a self-consistent static $GW$ approximation (COHSEX) calculation followed by a standard dynamic $G_0W_0$ last step of calculation (SC-COHSEX+$G_0W_0$ scheme);~\cite{bruneval.2006}
and, on the other side, the quasiparticle self-consistent $GW$ (QS$GW$) scheme.~\cite{vanschilfgaarde.2006,kotani.2007}
Both may improve the DFT wavefunctions and eigenvalues.

Hybrid functionals have also been proposed into the framework of an unrestricted DFT to solve the typical bandgap underestimation of the LDA and GGA approximations.
In these functionals, a fixed amount of Hartree-Fock exact non-local exchange is incorporated into the classical DFT local Kohn-Sham exchange-correlation potential.
Among these, the one proposed by Heyd, Scuseria, and Ernzerhof\cite{heyd.2003,heyd.2004,heyd.2005} (HSE) has been widely used lately.
Hybrid functionals have proven to perform well for improving several properties of solids.~\cite{heyd.2003,peralta.2006}
A typical HSE calculation is usually more computationally demanding than LDA or GGA, but more affordable than $GW$.

In this paper, the bandstructure of gold is calculated within \textit{ab initio} many-body perturbation theory (MBPT) in order to elucidate the role of correlations and to provide a more reliable theoretical bandstructure to interpret the experimental findings.
Standard $G_0W_0$ corrections shift the unoccupied bands up by at most 0.2~eV and the first $sp$ character occupied band down, while leaving the 5$d$ occupied bands unmodified.
Self-consistency on the wavefunctions by the QS$GW$ scheme lowers the 5$d$ bands by 0.4~eV, reducing the discrepancy with the experimental measurements.
Inclusion of $sp$ semicore states is confirmed to be crucial for $GW$ calculations in $d$-electron systems, as previously found.\cite{marini.2001,marini.2002}
In contrast, here the plasmon-pole model (PPM) is found to be overall valid.
The importance of relativistic effects in gold is also confirmed.~\cite{romaniello.2005}
The remaining disagreement with the experiments might be explained by the lack of relativistic many-body terms\cite{itoh.1965,olevano.2010,sakuma.2011} beyond the single-particle ones taken into account here.

Finally, we calculate the HSE hybrid functional bandstructure of gold and compare it to the QS$GW$ results.
Around the Fermi energy, HSE (and PBE) bands present a difference of $\sim$0.3 eV from the corresponding QS$GW$ ones.
High energy unoccupied HSE bands present a large discrepancy, by more than 6 eV, with respect to the experimental data and the $GW$ results.

The article is organized as follows.
In section~\ref{background}, the theoretical background is given.
The technical details of the calculations are shown in section~\ref{tec.dets}.
In section~\ref{g0w0-gold}, the bandstructure calculated within the $G_0W_0$ approach is analyzed.
The role of semicore orbitals and the validity of the PPM are discussed here.
In section~\ref{QSGW-gold}, the bandstructure calculated within the QS$GW$ method is presented.
Spin-orbit corrections are discussed in section~\ref{so-gold}.
In section~\ref{residual-discrepancies} we discuss the weight of all our approximations with respect to the residual discrepancies with the experiment.
An analysis of the HSE results is presented in section~\ref{hse-gold}. 
Finally, in section~\ref{conclusions-gold}, the conclusions of this work are drawn. 
In addition, convergence issues are discussed in the supplemental material.~\cite{supplemental}

\section{Theoretical background}\label{background}

In MBPT, the electronic structure is obtained by solving the quasiparticle (QP) equation:~\cite{hedin.1965,hybertsen.1985,hybertsen.1986,godby.1986,godby.1988,pulci.2005}
\begin{eqnarray}
\label{Eqn:many}
&&
\left( -\frac{1}{2}\nabla^2 + v^\mathrm{ext}(\mathbf{r}) + v^\mathrm{H}(\mathbf{r}) \right)
\psi_{n\mathbf{k}}^\mathrm{QP}(\mathbf{r}) + \nonumber \\
&&
\int d^{3}\mathbf{r'} \, \Sigma(\mathbf{r},\mathbf{r^\prime},\omega=
\epsilon_{n\mathbf{k}}^\mathrm{QP})\psi_{n\mathbf{k}}^\mathrm{QP}(\mathbf{r'})=
\epsilon_{n\mathbf{k}}^\mathrm{QP}\psi_{n\mathbf{k}}^\mathrm{QP}(\mathbf{r}),
\end{eqnarray}
where $v^\mathrm{ext}(\mathbf{r})$ is the external potential, $v^\mathrm{H}(\mathbf{r})$ is the classical repulsion Hartree term, and $\Sigma(\mathbf{r},\mathbf{r^\prime},\omega)$ is the self-energy, a non-hermitian, non-local and energy dependent operator.
The exact self-energy can be written as $\Sigma = G W \Gamma$, an expression containing the single particle Green's function $G$, the dynamically screened Coulomb potential $W$ and the vertex function $\Gamma$.
Hedin~\cite{hedin.1965} provided a scheme based on a closed set of five Schwinger-Dyson integro-differential equations for $G$, $W$, $\Gamma$, $\Sigma$ and the polarizability $P$ to be solved iteratively up to the self-consistent solution for $G$ and $\Sigma$.
Since the application of this scheme to real systems is usually computationally unfeasible, further approximations are required.
Setting $\Gamma=\delta$, the self-energy operator becomes
\begin{equation}
\label{Eqn:GW}
\Sigma(\mathbf{r,r^\prime},\omega)=
\frac{i}{2\pi}\int d\omega^\prime e^{i\omega^{\prime}\eta} G(\mathbf{r,r^\prime},\omega+\omega^\prime)
W(\mathbf{r,r^\prime},\omega^\prime),
\end{equation}
where $\eta$ is an infinitesimal positive number.
Due to its form, this is called the $GW$ approximation.
Starting from an initial approximation $G_0$ for the Green's function (for example, the one constructed from DFT orbitals), one can iterate the equations up to self-consistency.
Alternatively, one can stop at the first iteration obtaining the so called $G_0W_0$ approximation.

In practice, it is very efficient to get QP energies using perturbation theory with respect to the DFT electronic structure, {\it i.e.}\ treating as perturbation the difference between the self-energy operator and the DFT exchange-correlation potential, $\Sigma- v_\mathrm{xc}$. 
The DFT eigenvalues $\epsilon^{\mathrm{DFT}}_{n\mathbf{k}}$ and eigenstates $\psi_{n\mathbf{k}}^{\mathrm{DFT}}$ are used as a zeroth-order approximation for their quasiparticle counterparts.
Thus, the QP energy $\epsilon_{n\mathbf{k}}^\mathrm{QP}$ is calculated by adding to $\epsilon_{n\mathbf{k}}^{\mathrm{DFT}}$ the first-order perturbation correction:
\begin{equation}
\label{Eqn:MBPT}
\epsilon_{n\mathbf{k}}^\mathrm{QP}=\epsilon^{\mathrm{DFT}}_{n\mathbf{k}}
+ Z_{n\mathbf{k}} \langle \psi^{\mathrm{DFT}}_{n\mathbf{k}} | \Sigma(\omega=\epsilon^{\mathrm{DFT}}_{n\mathbf{k}})-v_{\mathrm{xc}} | \psi^{\mathrm{DFT}}_{n\mathbf{k}} \rangle,
\end{equation}
with $Z$ the quasiparticle renormalization factor, 
\begin{equation}
 Z= \left[ 1 - 
   \langle 
\psi^\mathrm{DFT}_{n\mathbf{k}} | 
   \left. 
\frac{\partial \Sigma(\omega)}{\partial \omega} 
\right|_{\omega=\epsilon^\mathrm{DFT}_{n\mathbf{k}}} | 
\psi^\mathrm{DFT}_{n\mathbf{k}}  
\rangle 
\right]^{-1}
,
\end{equation}
which accounts for the fact that, in Eq.~(\ref{Eqn:many}), $\Sigma(\omega)$ should be calculated at the $\epsilon_{n\mathbf{k}}^\mathrm{QP}$. 
This procedure has been found to produce bandstructures in agreement with the experiment, provided that the DFT states are not too far from the QP states. Otherwise, a self-consistent approach on the eigenvalues and eigenstates may be necessary.

In the so-called QS$GW$ calculations,\cite{vanschilfgaarde.2006,kotani.2007} the self-energy is constrained to be Hermitian and static, so that it can be diagonalized to update not only the energies, but also the wavefunctions.
Several successive iterations are needed to achieve the desired accuracy.
At the end, the self-energy does not depend anymore on the DFT starting point.

The integration of Eq.~(\ref{Eqn:GW}) requires in principle the evaluation of $W(\omega)$ over a large number of frequencies.
By modeling $\Im W(\omega)$ with a single pole in the plasmon-pole model (PPM),~\cite{hybertsen.1986,godby.1989} it is possible to integrate Eq.~(\ref{Eqn:GW}) analytically.
In the case of $d$ electrons, the applicability of this technique has been questioned.~\cite{marini.2001}
More accurate integration methods, such as the contour deformation (CD) approach, are frequently used.
In this technique, the real axis integration path of Eq.~(\ref{Eqn:GW}) is modified as to run along the imaginary axis, picking up contributions coming from the poles of the Green's function included in the deformed contour.~\cite{aryasetiawan.2000,kotani.2002,lebegue.2003}

In principle, to fully take into account single-particle relativistic effects, one should solve the Dirac equation and work with Dirac spinors.
Alternatively, one can use a non-relativistic limit of the Dirac equation projected onto a Pauli two-component spinor formalism.
This adds the fine structure terms to the Hamiltonian.
In the standard limit approach, there are three such terms: the $p^4$ relativistic correction to the velocity, the Darwin term, and the spin-orbit (SO) coupling.
The scalar-relativistic approach includes only the first two terms and drops the SO coupling term.
In some cases, the resulting equation already accounts for most of the Dirac physics.
If needed, the SO coupling effects can be introduced on top of the scalar-relativistic approach, using the procedure detailed in Sec.~\ref{so-gold}.
However, in the most severe cases, the SO coupling effects should be introduced from the beginning in a fully spinorial formalism.~\cite{aryasetiawan.2008,aryasetiawan.2009} 
So far, this formalism has only been applied to the bandstructure of Hg chemical compounds,~\cite{sakuma.2011} finding SO coupling corrections to the eigenvalues of $\sim$0.1~eV.
This calculation was carried on only up to the first iteration of Hedin's equations, {\it i.e.}\ at the $G_0W_0$ level.
Going further in the direction of self-consistency and including relativistic corrections, has not yet been tried on any real system.

In the case of gold, most of the relativistic effects in the bandstructure come from the scalar-relativistic terms.~\cite{christensen.1971,romaniello.2005}
The SO coupling term mainly accounts for band splittings, hence, it introduces shape modifications mostly on the 5$d$ bands.~\cite{christensen.1971,romaniello.2005}

\section{Technical details}
\label{tec.dets}

All calculations are performed using the primitive unit cell of gold (FCC lattice). 
Note that in principle van der Waals interactions are important to determine the atomic distance in noble metals.~\cite{maggs.1987}
To avoid this difficulty the experimental lattice constant (7.71~Bohr~\cite{ashcroft.1966}) is used.~\cite{note:lattice}
The $GW$ calculations are done using the \textsc{Abinit} code,~\cite{abinit} while the HSE ones are carried out with the \textsc{Vasp} code.~\cite{vasp}
Scalar relativistic effects have been included everywhere.

In the $GW$ calculations, the starting point wavefunctions and energies are obtained from a DFT calculation in which the XC energy is approximated by the GGA PBE functional.~\cite{perdew.1996}
Scalar-relativistic norm-conserving pseudopotentials~\cite{teter.1993,grinberg.2000} are used to account for core-valence interactions.~\cite{note:pseudo}
In order to elucidate the role of semicore states, two pseudopotentials are considered.
The first one contains 11 valence electrons (5$d^{10}$, 6$s^{1}$), while the second contains 19 electrons (5$s^2$, 5$p^6$, 5$d^{10}$, 6$s^1$).
The wavefunctions are expanded on plane-waves basis sets, up to a cut-off energy of 30~Ha when the semicore states are not included, and 50~Ha when they are.
The Brillouin zone (BZ) is sampled using a shifted grid of 10$\times$10$\times$10 $k$-points following the Monkhorst-Pack (MP) scheme.~\cite{monkhorst.1976}
A total of 110 (100 empty) bands are used to compute the dielectric matrix\cite{botti.2002} and the self-energy.
The dielectric matrix is computed for 145 $k$-points in the irreducible BZ, truncating to an energy cut-off of 4.0 Ha (corresponding to 59 plane waves).
The Godby-Needs PPM~\cite{godby.1989} is used here because it has demonstrated the best agreement with the methods which take fully into account the frequency dependence of the dielectric matrix.~\cite{shaltaf.2008,stankovski.2011}
In the CD method, a total of 6 and 20 frequencies are used along the imaginary and real axis, respectively.
All QS$GW$ calculations are performed within the CD method.
A total of 40 bands are considered when diagonalizing the self-energy.

In the calculations with the hybrid XC functional, only 11 valence electrons are treated explicitly by the projector augmented-wave (PAW) method.
The plane-wave cut-off energy for the wavefunctions is chosen to be 13~Ha.
HF type calculations are performed with the HSE06 functional~\cite{heyd.2003}, starting from previously converged DFT wavefunctions and energies.
These calculations are considerably more costly than standard DFT ones.
Hence, we could only afford to sample the BZ using a 20$\times$20$\times$20 unshifted $\Gamma$ grid of $k$-points.

In all cases, the bandstructures are interpolated using maximally-localized Wannier functions (MLWFs) with the Wannier90 code~\cite{wannier90} as explained in Refs.~\onlinecite{yates.2007,hamann.2009}.
The Fermi level is obtained by integrating the density of states (DOS), calculated with an interpolated grid of 30$\times$30$\times$30 $k$-points using MLWFs and a low Gaussian smearing of 0.005~Ha.
It was verified that the Fermi levels obtained with a grid of 30$\times$30$\times$30 and
60$\times$60$\times$60 interpolated $k$-points were equal within 0.01~eV.

A full study of the convergence with respect to all parameters of the calculation is provided in the supplemental material.~\cite{supplemental}

\begin{figure}
\includegraphics{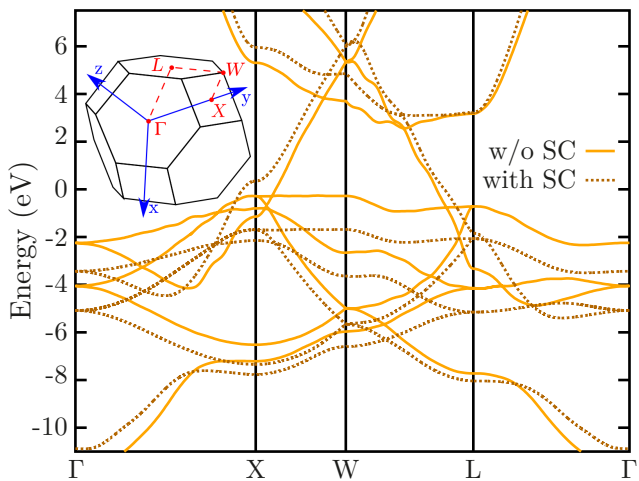}
\caption{(Color online) Effect of the semicore orbitals on the bandstructure of gold calculated within $G_0W_0$ using a plasmon-pole model. The results obtained when the semicore states are not considered as valence electrons (w/o SC) are represented by solid orange (light grey) lines, while those calculated with the semicore states treated as valence electrons (with SC) are shown as dotted brown (medium grey) lines.
The zero of energy has been set at the Fermi level.
The corresponding Brillouin zone is shown on top.
All the calculations in this paper are performed at least at the scalar-relativistic level.
\label{fig:g0w0-ppm}}
\end{figure}

\section{The $G_0W_0$ bandstructure of gold}
\label{g0w0-gold}

In this section, we investigate the QP bandstructure of gold within the $G_0W_0$ approach, trying to clarify the influence of two commonly used approximations.
First, the effect of freezing semicore orbitals in the pseudopotential is discussed.
Second, the validity of the PPM is analyzed more thoroughly.

In Fig.~\ref{fig:g0w0-ppm}, the bandstructure of gold calculated within $G_0W_0$ is reported using two different pseudopotentials. In the first one [solid orange (light grey) lines, labeled ``w/o SC"], the $5s$ and $5p$ semicore orbitals are considered to be frozen in the core (leading to a total of 11 valence electrons). In the second one [dotted brown (medium grey) lines, labeled ``with SC"], 19 electrons are treated as valence states. While within DFT the resulting bandstructures are on top of each other (the curves are not shown here for sake of clarity), the difference becomes important at the $GW$ level. Indeed, when the semicore electrons are excluded (``w/o SC"), the $5d$ bands are shifted up while the $6sp$ bands are shifted down in a non-homogeneous way.
This leads to a reduction of the $5d$-$6sp$ interband gap.
This effect is alarming in the neighborhood of the X point, where the lowest empty band is shifted by -1.7~eV while the top-most 5$d$ band is shifted by +1.1~eV, thus leading to an inversion in the band ordering.
This unphysical shifting of bands is solved by including the exchange contributions from the $5d$ to the $5s$ and $5p$ semicore orbitals (``with SC").
Although $5s$ and $5p$ states are separated in energy by more than 50~eV from the $5d$ ones, their spatial overlap with the $5d$ is important.
Hence, they play an important role at the $GW$ level and cannot be neglected.\cite{rohlfing.1995,marini.2001} 
In the remainder of the paper all the $GW$ calculations are performed treating explicitly these electrons as valence states.

\begin{figure}
\includegraphics{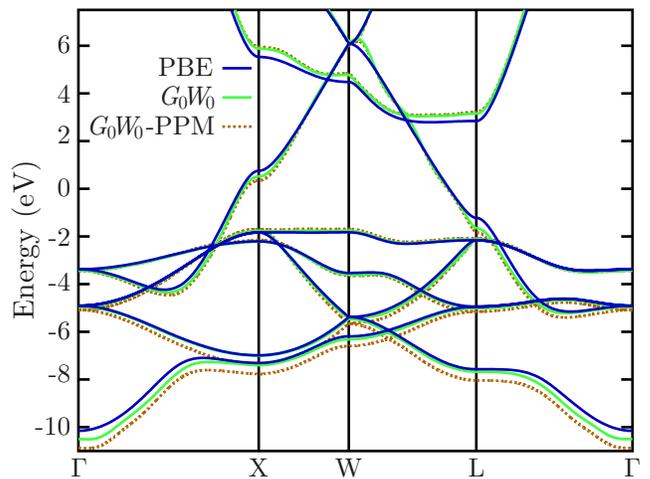}
\caption{(Color online) Bandstructure of gold calculated within DFT-PBE [solid blue (black) lines] and $G_0W_0$ using the contour deformation technique [solid green (light grey) lines] or the Godby-Needs plasmon-pole model [dotted brown (medium grey) lines].
The zero of energy has been set at the Fermi level.
\label{fig:g0w0}}
\end{figure}

PPMs are believed not to work satisfactorily in the presence of $d$-electrons just below the Fermi level.
Indeed, this may induce strong transitions in $\epsilon^{-1}_{\mathbf{GG'}}(\mathbf{q},\omega)$.
As a result, this function cannot always be approximated by a single-pole function at small values of $G$ and $G'$.~\cite{marini.2001}
Fig.~\ref{fig:g0w0} shows the bandstructure of gold calculated within $G_0W_0$ using either a PPM [dotted brown (medium grey) lines] or the more accurate CD method [solid green (light grey) lines].
For bands located in the energy window going from the Fermi level to 5~eV below, both methods give similar results (within a maximum difference of 0.1~eV).
Below this window, the use of the PPM tends to shift the bands down compared to CD, with a discrepancy which can be up to 0.2~eV.
This PPM inaccuracy on the lowest band is also present in other systems,
such as in silicon and diamond,~\cite{hybertsen.1985} whose energy-loss function~(ELF) presents a well-defined single plasmon resonance.~\cite{olevano.2001}
Although in noble metals the ELF has a more complex structure, the single PPM cannot be considered less valid in gold than in silicon and diamond.
In what follows we will anyway use the CD method for all $GW$ calculations.

In Fig.~\ref{fig:g0w0}, the DFT-PBE bandstructure of gold [solid blue (black) lines] is also reported.
It is found to be in agreement with previous calculations.~\cite{romaniello.2005}
The $G_0W_0$ bandstructure [solid green (light grey) lines] is almost on top of the DFT-PBE one, but the first unoccupied band is shifted up non-homogeneously by up to $\sim$0.2~eV and the first occupied band is shifted down by $\sim$0.4~eV at $\Gamma$. 
These bands present a predominant $sp$ character.
The $G_0W_0$ corrections are, anyway, not modifying the $5d$ manifold of bands:
their shape, position and bandwidths are the same as in the DFT-PBE case.
As a consequence, the $G_0W_0$ 5$d$-6$sp$ interband gap does not change compared to the DFT-PBE value, which is smaller than the experimental evidence.

\begin{figure}
\includegraphics{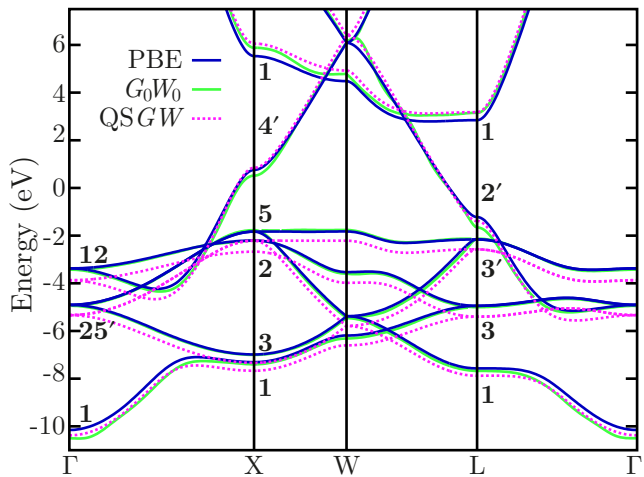}
\caption{(Color online)
Bandstructure of gold calculated within DFT-PBE [solid blue (black) lines], $G_0W_0$ [solid green (light grey) lines], and QS$GW$ [dotted pink (grey) lines].
All $GW$ calculations are done within the CD method.
The zero of energy has been set at the Fermi level.}
\label{scgw}
\end{figure}

\section{Self-consistency effects within the QS$GW$ approximation}
\label{QSGW-gold}

Fig.~\ref{scgw} shows the bandstructure for different approaches: DFT-PBE [solid blue (black) lines], $G_0W_0$ [solid green (light grey) lines] and QS$GW$ [dotted pink (grey) lines].
The transition energies at high symmetry $k$-points can also be read in Table~\ref{transitions}.

\begin{table}
\begin{ruledtabular}
\begin{tabular}{clcccc}
& & PBE & $G_0W_0$ & QS$GW$\\
\hline
&$\Gamma_1$~$\rightarrow$~$\Gamma_{25'}$  & 5.2 & 5.6 & 5.0\\
& $\Gamma_{25'}$~$\rightarrow$~$\Gamma_{12}$& 1.5 & 1.5 & 1.5\\
&X$_3$~$\rightarrow$~X$_2$    & 4.8 & 4.8 & 4.7\\
&X$_5$~$\rightarrow$~X$_{4'}$ & 2.6 & 2.3 & 3.1\\
&X$_{4'}$~$\rightarrow$~X$_1$ & 4.8 & 5.4 & 5.2\\
&L$_3$~$\rightarrow$~L$_{3'}$ & 2.8 & 2.9 & 2.8\\
&L$_{3'}$~$\rightarrow$~L$_{2'}$ & 1.0 & 0.4 & 1.2\\
&L$_{2}$~$\rightarrow$~L$_1$ & 4.0 & 4.8 & 4.6
\end{tabular}
\end{ruledtabular}
\caption{Transition energies of gold (in eV) calculated within scalar-relativistic DFT-PBE, $G_0W_0$, and QS$GW$.}
\label{transitions}
\end{table}

When recalculating the QP wavefunctions within the QS$GW$ approach, the $5d$ bands are shifted with respect to DFT-PBE by $-0.4$~eV.
This is the major difference with respect to one shot $G_0W_0$.
In addition, the first unoccupied bands are further shifted, achieving $+0.3$~eV from DFT-PBE.
As a consequence, the interband gap between the 5$d$ and the unoccupied bands is opened by 0.4 to 0.8~eV with respect to the DFT-PBE energies.
For instance, the transition energies X$_5$~$\rightarrow$~X$_{4'}$ and L$_{3'}$~$\rightarrow$~L$_1$ are opened by 0.45 and 0.75~eV, respectively.
This points out to the significance of correcting the DFT-PBE wavefunctions in order to obtain a more accurate bandstructure.

\begin{figure}
\includegraphics{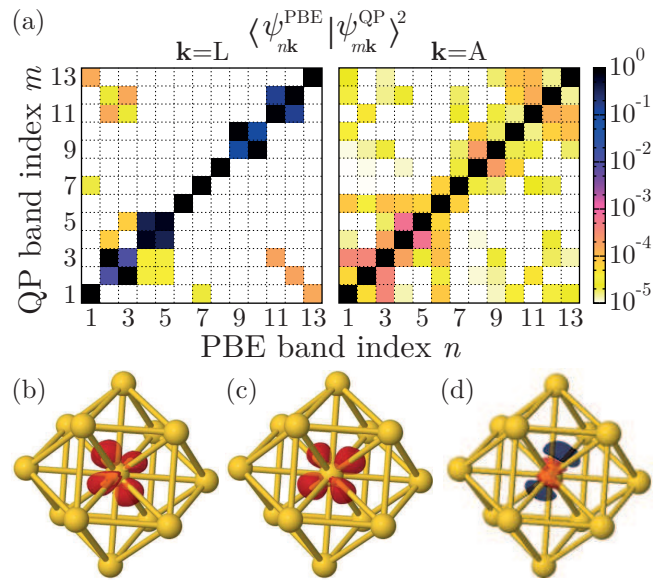}
\caption{(Color online)
Illustration of the DFT band mixing at the QS$GW$ level.
Panel (a) represents the square of the overlap between the QP and DFT-PBE wavefunctions at the L (left) and A (right) $k$-points.
The A point is a random low symmetry $k$-point with reduced coordinates (0.5,0.3,0.1).
The square modulus of the QP and DFT-PBE wavefunctions, $|\psi_{n\mathbf{k}}^\mathrm{QP}|^2$ and $|\psi_{n\mathbf{k}}^\mathrm{DFT}|^2$, for band index $n$=4 at $k$-point L are shown in panels (b) and (c), respectively.
Panel (d) shows the difference $|\psi_{n\mathbf{k}}^\mathrm{QP}|^2-|\psi_{n\mathbf{k}}^\mathrm{DFT}|^2$ for band index $n$=6 at $k$-point A.
Gold atoms in the FCC lattice are represented by yellow [light grey] spheres.
In panels (b)-(d), the isosurfaces correspond to $+1\rho$ in red (grey) and $-1\rho$ in blue (black), with $\rho= 6\times10^{-4} e^{\textrm{-}} $/\AA$^3$ for panels (b) and (c), while for panel (d), $\rho= 3\times10^{-5} e^{\textrm{-}} $/\AA$^3$.
}
\label{dft-qps}
\end{figure}

To understand the effect of quasiparticle self-consistency, the QP and DFT-PBE wavefunctions are compared in Fig.~\ref{dft-qps}.
It is found that QS$GW$ introduces a mixing of DFT-PBE states which corresponds to rotations and small relocalizations of the wavefunctions.
These changes depend on the $k$-point $\mathbf{k}$ and the band index $n$.

In Fig.~\ref{dft-qps}(a), we plot the square of overlap between the QP and DFT-PBE wavefunctions at $k$-points L and A, the latter being a random low symmetry $k$-point with reduced coordinates (0.5, 0.3, 0.1).
This is a direct indication of the band mixing resulting from the QS$GW$ procedure.
The square modulus of the QP and DFT-PBE wavefunctions, $|\psi_{n\mathbf{k}}^\mathrm{QP}|^2$ and $|\psi_{n\mathbf{k}}^\mathrm{DFT}|^2$, for band index $n$=4 at the L point are shown in Fig.~\ref{dft-qps}(b), and panel (c) respectively.
Finally, in Fig.~\ref{dft-qps}(d), we report the difference $|\psi_{n\mathbf{k}}^\mathrm{QP}|^2-|\psi_{n\mathbf{k}}^\mathrm{DFT}|^2$ for band index $n$=6 at the A point.
Regardless of the $k$-point, the strongest mixing is always found between degenerate bands (see top panel).
It gives rise to rotations of the wavefunctions associated to individual bands.
For example, the QP wavefunction associated to band index $n$=4 at the L point [Fig.~\ref{dft-qps}(c)] corresponds simply to a spatial rotation of the corresponding DFT-PBE wavefunction [Fig.~\ref{dft-qps}(b)] around the center of a gold atom.
In fact, bands $n$=4 and 5 are degenerate in energy and the corresponding wavefunctions have the same symmetry with a different orientation. Therefore, the mixing of these bands just induces a change in the orientation of the wavefunctions.
The wavefunctions associated to other degenerate bands may also undergo similar rotations, without any noticeable effect on the bandstructure.~\cite{note:mixing}

More importantly, numerous small hybridizations occur between the occupied bands and the higher empty bands [Fig.~\ref{dft-qps}(a)].
This is more evident at low-symmetry $k$-points, such as the A point.
These small hybridizations may have an important effect on the shape and localization of the wavefunctions.
To illustrate this, we calculate the difference between the square modulus of the QP and DFT-PBE wavefunctions.
This is done for the first unoccupied band at A.
For this particular band and $k$-point, a relocalization of the wavefunction is observed:
the 5$d$ character is reduced [blue (black) lobes] while the 6$s$ character close to the atom is slightly augmented [red (grey) lobes].
The nature of these changes depends on the $k$-point and the band index $n$.
The effect of these changes of the wavefunctions is that the diagonal elements of the self-energy $\langle \psi_{n\mathbf{k}} | \Sigma | \psi_{n\mathbf{k}} \rangle$ and Hartree $\langle \psi_{n\mathbf{k}} | v^\mathrm{H} | \psi_{n\mathbf{k}} \rangle$ operators are modified, inducing an almost rigid shift of about 0.4~eV downward of the 5$d$ bands.

\begin{figure}
\includegraphics{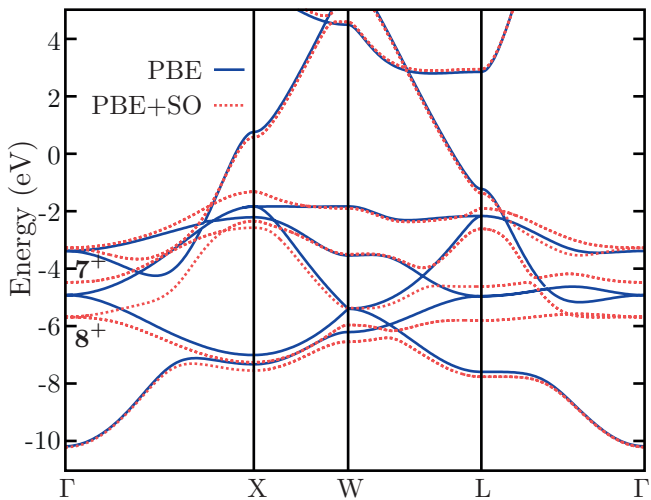}
\caption{
(Color online) DFT-PBE bandstructure of gold calculated within the scalar-relativistic (SR) approximation [solid blue (black) lines] and including also the spin-orbit coupling (SR+SO) [dotted red (grey) lines].
The zero of energy is set at the Fermi level.}
\label{fig:bandsso}
\end{figure}

\section{Spin orbit coupling effects}
\label{so-gold}

In order to fully take into account relativistic effects at least at the single-particle level, in principle one should solve the Dirac equation and work with Dirac spinors.
Alternatively, one can continue to work with Pauli spinors by choosing an appropriate non-relativistic limit of the Dirac equation which adds some relativistic corrections to the Schr\"odinger equation Hamiltonian.
In the scalar-relativistic (SR) approximation, one solves a Schr\"odinger equation including the relativistic correction to the velocity by the mass and the Darwin terms.
These terms may cause important band-shifts and they should already capture most of the relativistic effect.~\cite{koelling.1977,takeda.1978,gollisch.1978}
In addition, one can include the spin-orbit (SO) coupling term which may cause important band-splitting and changes to the band-shape.
Hereafter, this procedure is referred to as SR+SO.

\begin{figure}
\includegraphics{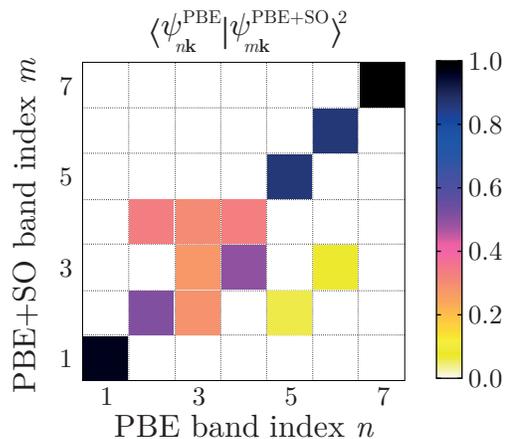}
\caption[SO effects on gold .]{
Square of the overlap between the scalar-relativistic (SR) and full relativistic (SR+SO) DFT-PBE wavefunctions.
}
\label{fig:proj-so}
\end{figure}

In Fig.~\ref{fig:bandsso} we show the comparison between the bandplot of a DFT-PBE calculation which only includes the SR terms in the Kohn-Sham Hamiltonian with that one of a fully relativistic (SR+SO) DFT-PBE calculation, which also includes the SO coupling.
In the case of gold, most of the relativistic effects in the bandstructure come from the scalar-relativistic terms.~\cite{christensen.1971,romaniello.2005}
The SO coupling term mainly accounts for band splittings, as shown in Fig.~\ref{fig:bandsso}.
To illustrate the effect of the SO coupling on the wavefunctions, the overlap between the SR and SR+SO DFT-PBE wavefunctions is calculated at the $\Gamma$ point, as shown in Fig.~\ref{fig:proj-so}.
The overlap is close to 1 for the occupied bands 1, 5 and 6, meaning that these bands are almost unaffected by the SO coupling term.
However, the $d$ bands 2, 3 and 4 are strongly changed by the SO coupling term.
The $\Gamma_{25'}$ state found in the scalar-relativistic calculation is split into the $\Gamma_{7^+}$ and $\Gamma_{8^+}$ states, once the SO coupling term is taken into account.
Similar effects are observed in other $k$-points as explained in Ref.~\onlinecite{romaniello.2005}.

Within MBPT, relativistic fine structure effects should in principle be calculated within a fully spinorial GW formalism.\cite{aryasetiawan.2008,aryasetiawan.2009}
So far, this formalism has been applied only to Hg compounds\cite{sakuma.2011} at the $G_0W_0$ level.
However, at the self-consistent level, this method has not yet been applied to real systems.

In this work, we add SO effects perturbatively on top of the QS$GW$ and HSE bandstructures by the following procedure:
\begin{enumerate}
\item We evaluate the SO corrections to DFT-PBE eigenvalues by a fully spinorial Kohn-Sham calculation;
\item We compute $\Sigma_{n\mathbf{k}}^\mathrm{SO}=\epsilon_{n\mathbf{k}}^\mathrm{SR+SO}-\epsilon_{n\mathbf{k}}^\mathrm{SR}$, the difference between the SR and SR+SO DFT-PBE eigenvalues at a given $k$-point and band index $n$.
\item We add $\Sigma_{n\mathbf{k}}^\mathrm{SO}$ to the corresponding QP (HSE) eigenvalue.
\end{enumerate}

\begin{figure}
\includegraphics{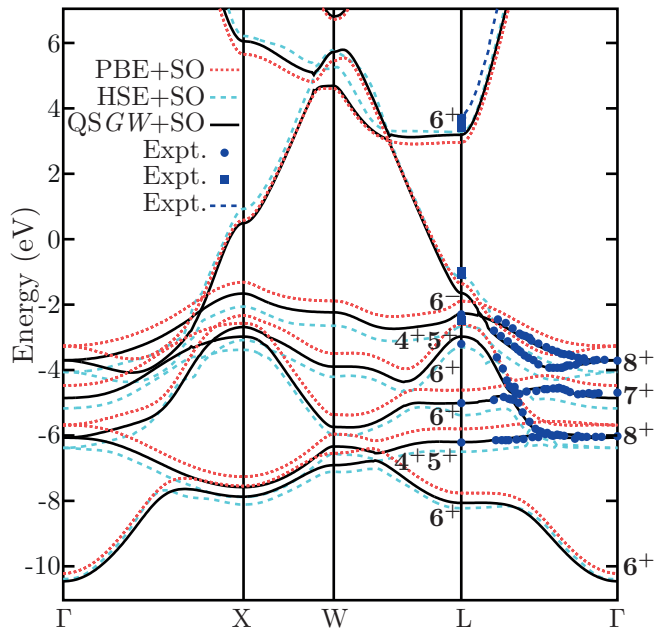}
\caption{(Color online) Bandstructure of gold calculated within PBE+SO [dotted red (grey) lines], QS$GW$+SO [solid black lines] and HSE+SO [dashed green (light grey) lines].
The zero of energy has been set at the Fermi level.
These theoretical results are compared to various experimental measurements.
The blue (black) circles are taken from Ref.~\onlinecite{courths.1986}.
At the L point, the blue (black) squares correspond to the measurements listed in Table~\ref{eigenvalues}.
The dashed blue (black) line gives the experimental final-band consistent with all data points from angle-resolved ultraviolet photoelectron spectroscopy~(ARUPS) in Refs.~\onlinecite{courths.1986,mills.1980,baalmann.1983} and from low-electron reflectance in Ref.~\onlinecite{jaklevic.1982}.
}
\label{bands-so}
\end{figure}

Fig.~\ref{bands-so} shows the PBE+SO [dotted red (grey) lines], QS$GW$+SO [solid black lines], and HSE+SO [dashed green (light-grey) lines] bandstructures including SO coupling effects.
The experimental bandstructure along the L~$\rightarrow~\Gamma$ $k$-path taken from Ref.~\onlinecite{courths.1986} is also shown.
The experimental and theoretical eigenvalues are listed in Table~\ref{eigenvalues}.

The QP occupied bands are in good agreement with the available experimental measurements with an average difference of 0.06~eV.
In fact, the 5$d$ bands are shifted by $-0.4$~eV, improving the agreement with the experimental data.
Indeed, this shift has been suggested before in Refs.~\onlinecite{heimann.1977,courths.1986}.
Nevertheless, the occupied L$_6^-$ band is lowered by 0.26~eV with respect to the DFT-PBE value, in the wrong direction with respect to the experiment [this is also the case in the bandstructures obtained within one-shot $GW$ (see Fig.~\ref{fig:g0w0})].
A disagreement of up to 0.6~eV in the first unoccupied band still remains.
To illustrate, for band 7 the discrepancy is of 0.4 and 0.6~eV at L$_6^+$ and $\Gamma_7^-$ (see Table~\ref{eigenvalues}).
Moreover, for higher energy bands, such as $\Gamma_6^-$ at 18 eV above the Fermi level, the deviations from the experimental data can be as large as~0.8~eV.

\section{Residual discrepancies}
\label{residual-discrepancies}

The inclusion into the QP bandstructure of spin-orbit effects by the present perturbative treatment might be considered as the source of the residual non-negligible discrepancies.
However, a more correct treatment within $GW$ of such effects, as in Ref.~\onlinecite{sakuma.2011}, was found to affect the result by not more than 0.1 eV.

The error due to the use of QS$GW$ instead of a full $GW$ self-consistency is presently unknown.
However, the use of a different self-consistent scheme, namely SC-COHSEX+$G_0W_0$, seems to provide results in agreement with QS$GW$.\cite{bruneval.2006}
Of course, one cannot exclude that both schemes at the same time provide deviations from full self-consistent $GW$ larger than 0.1 eV.

Other possible sources of these discrepancies might be vertex corrections beyond $GW$.
Here we checked the local vertex correction\cite{delsole.2003} and a non-local vertex correction to $W$ only\cite{stankovski-nonloc}.
These account for small corrections of no more than 0.1~eV, as explained in Ref.~\onlinecite{rangel.2011}.

Intraband $q\to 0$ Drude peak contributions to the polarizability, which were neglected in our calculations, may lead to a spurious gap at the Fermi level in simple (alkali) metals\cite{cazzaniga}.
However, no spurious gaps were observed here.
In fact, the neglect of the Drude peak in slightly more complex metals, such as aluminium, does not lead to significant errors.\cite{cazzaniga}

The relativistic corrections taken into account here, as well as in Ref.~\onlinecite{sakuma.2011}, are only at the single-particle level.
At present, the effect of many-body relativistic terms\cite{itoh.1965,olevano.2010}, such as the Breit interaction or the spin-of-one-electron orbit-of-the-second\cite{itoh.1965}, etc., is unknown.
In systems like gold, where relativistic effects are important, these terms might explain the remaining discrepancies.

\begin{table}
\begin{ruledtabular}
\begin{tabular}{lcrrrc}
\multicolumn{2}{c}{Symmetry} &
\multicolumn{1}{c}{PBE} &
\multicolumn{1}{c}{QS$GW$} &
\multicolumn{1}{c}{HSE} & Expt.\\
\multicolumn{2}{c}{label}&
\multicolumn{1}{c}{+SO} &
\multicolumn{1}{c}{+SO} &
\multicolumn{1}{c}{+SO}  \\
\multicolumn{2}{c}{(band index) } \\
\hline
$\Gamma^+_6$ &(1) &-10.19 &-10.39&-10.30&\\
$\Gamma^+_8$ &(2,3) & -5.67& -6.02 & -6.31 &
-5.90\footnote{Angle resolved ultraviolet photoelectron spectroscopy (Ref.~\onlinecite{mills.1980})},%
-6.0\footnote{Angle resolved ultraviolet photoelectron spectroscopy (Ref.~\onlinecite{baalmann.1983})},%
-6.01 (0.02)\footnote{Angle resolved ultraviolet photoelectron spectroscopy (Ref.~\onlinecite{courths.1986})}\\
$\Gamma^+_7$ &(4) & -4.46 & -4.85 & -4.82 &-4.45\footnotemark[1],%
-4.6\footnotemark[2],-4.68 (0.05)\footnotemark[3]\\
$\Gamma^+_8$ &(5,6) & -3.27& -3.67 & -4.00 &-3.55\footnotemark[1],%
-3.65\footnotemark[2],-3.71 (0.02)\footnotemark[3]\\
$\Gamma^-_7$ &(7) & 15.76 & 15.36& 23.27&
16.0 (0.1)\footnotemark[3],%
15.9\footnote{Low-energy-electron reflectance (Ref.~\onlinecite{jaklevic.1982})}\\
$\Gamma^-_6$ &(8) & 18.08& 17.97& 24.38&
18.8 (0.5)\footnotemark[3]\\
L$^+_6$ &(1) & -7.74& -8.01 & -8.15&
-7.80 (0.15)\footnotemark[2]\\
L$^+_{4,5}$ &(2) & -5.79& -6.16 & -6.40&
-6.23 (0.15)\footnotemark[2],%
-6.20 (0.05)\footnotemark[3]\\
L$^+_6$                &(3) & -4.61& -4.97 & -5.36&
-4.88 (0.1)\footnotemark[2],%
-5.0 (0.05)\footnotemark[3]\\
L$^+_6$                &(4) & -2.61& -2.95 & -3.25&
-3.2 (0.1)\footnotemark[3]\\
L$^+_{4,5}$ &(5) & -1.90& -2.24 & -2.60&
-2.3 (0.1)\footnotemark[3],%
-2.5\footnote{Piezoreflectance~\cite{szczepanek.1975}}\\
L$^-_6$              &(6) & -1.37& -1.63 & -1.12&-1.0\footnotemark[5],%
-1.0 (0.1)\footnote{Electro tunneling (Ref.~\onlinecite{jaklevic.1975})}, \\
&&&&&
-1.01 (0.04)\footnote{Piezooptical response (Ref.~\onlinecite{chen.1976})},%
-1.1 (0.1)\footnote{Bremsstrahlung isochromat spectroscopy (Ref.~\onlinecite{marel.1984})}\\
L$^+_6$              &(7) & 2.93& 3.19 & 3.29&
3.6\footnotemark[5],%
3.65 (0.05)\footnotemark[6],\\
&&&&&
3.56 (0.02)\footnotemark[7],%
3.4 (0.1)\footnotemark[8]\\
\end{tabular}
\end{ruledtabular}
\caption{Experimental and theoretical values (in eV) for the energy bands of gold at the high-symmetry points $\Gamma$ and L.
The theoretical results include SO coupling corrections (see the text). 
Experimental errors are shown in parentheses (eV).}
\label{eigenvalues}
\end{table}

\section{The HSE bandstructure of gold}\label{hse-gold}

Within HSE, the partially-occupied bands close to the Fermi level are in good agreement with the QP and experimental energies.
For instance, the position of L$^-_6$ is within 0.1~eV of the experimental data (see Table~\ref{eigenvalues}).
For this particular point, HSE presents a better agreement with the experimental data than QS$GW$ does.
The QP and HSE bands along the W to X and $\Gamma$ to L paths
agree almost perfectly from -1 to 3~eV [the Fermi level is at zero] (see Fig.~\ref{bands-so}).
However, in this energy range, a disagreement of~$\sim$0.4 eV is found
 in the vicinity of the X point.
Moreover, the HSE 5$d$ bands are $\sim$0.3~eV below the QS$GW$ results and the experimental data.
This shows that HSE opens the interband gap between the unoccupied and the 5$d$ bands too much .
For higher energy bands, the agreement is quite poor.
For instance, the HSE eigenvalues at the $\Gamma^-_7$ and $\Gamma^-_6$ points are $\sim$6 to 7~eV above the $GW$ and experimental data.

Our findings, and in contemporary those of other authors~\cite{jain.2011}, show that the HSE functional does not systematically predict reliable band widths and gaps. 
In fact, the amount of exact exchange in the HSE functional is chosen so to provide good structural, thermochemical and bonding properties of solids~\cite{paier.2006,marsman.2008}.
For metals, our results, in agreement with Refs.~\onlinecite{paier.2006,marsman.2008}, show that HSE overestimate transition energies. 
Moreover, the modification in the $d$ wavefunctions as provided by self-consistent $GW$ are not catched by HSE, and the corresponding physics is not reproduced.

\section{Conclusions}
\label{conclusions-gold}
In summary, we have studied the bandstructure of gold using MBPT with several flavors of the $GW$ approximation and using the HSE hybrid functional.
While the inclusion of semicore 5$s$ and 5$p$ states in the valence shell has negligible effects in DFT, it becomes crucial in $GW$, leading to a wrong inverse ordering of bands at the Fermi level when they are neglected.
Within $G_0W_0$, the plasmon-pole model is found to be a good approximation for gold.
The PPM provides the same results, within 0.1~eV, as the full contour-deformation integration method, except for the lowest bands where deviations can be up to 0.2~eV.
With respect to DFT-PBE, the single-shot $G_0W_0$ shifts the empty bands up by $\sim$0.2~eV and the lowest $sp$ occupied band down by 0.4~eV, while leaving the 5$d$ occupied bands unchanged.
Updating the DFT-PBE wavefunctions, as in the QS$GW$ approach, is important to shift down by 0.4~eV the occupied 5$d$ bands, thus improving the agreement with the experiment.
A residual discrepancy of up to 0.6~eV in the 5$d$-6$sp$ interband gap is still present, probably due to relativistic effects beyond those included here, as well as, the lack of a unified relativistic many-body approach.
Finally, the position of the 5$d$ bands calculated within HSE ends up $\sim$0.3~eV below the experimental data.
HSE becomes more and more off for higher states, with an error of $\sim$6~eV at 16~eV from the Fermi level.

\begin{acknowledgments}
We thank Martin Stankovski for his valuable comments and interesting discussions.
This work was supported by the EU FP6 and FP7 through the Nanoquanta NoE (NMP4-CT-2004-50019) and the ETSF I3 e-Infrastructure (Grant Agreement 211956), and the project FRFC N$^\circ$. 2.4502.05.
We thank the the French Community of Belgium for financial support via the Concerted Research Action programme (ARC NANHYMO: convention 07/12-003).
DK and HVS acknowledge the financial support from the Competence Center for Materials Science and Technology (CCMX) and the computing facilities at CSCS, both in Switzerland.
PET thanks the financial support from the ERC Starting Grant FP7 Project DEDOM (No. 207441).
Most of the computer time has been provided by the supercomputing facilities of the Universit\'e catholique de Louvain (CISM/UCL), by the Consortium des Equipements de Calcul Intensif en F\'ed\'eration Wallonie Bruxelles (CECI) funded by the Fond de la Recherche Scientifique de Belgique (FRS-FNRS), and by the Ciment/Phynum centre via the {\it Fondation Nanosciences} NanoSTAR project.

\end{acknowledgments}

\bibliography{basename of .bib file}

\begin{thebibliography}{98}

\bibitem{christensen.1971}
N. E. Christensen and B. O. Seraphin,
Phys. Rev. B {\bf 4}, 3321 (1971).

\bibitem{heimann.1977}
P. Heimann and H. Neddermeyer,
J. Phys. F {\bf 7}, L37 (1977).


\bibitem{pyykko.1979}
P. Pyykko and J. P. Desclaux,
Acc. Chem. Res. {\bf 12}, 276 (1979).



\bibitem{maggs.1987}
A. C. Maggs and N. W. Ashcroft,
Phys. Rev. Lett. {\bf 59}, 113 (1987).

\bibitem{courths.1986}
R. Courths, H. G. Zimmer, A. Goldmann, and H. Saalfeld,
Phys. Rev. B {\bf 34}, 3577 (1986).

\bibitem{romaniello.2005}
P. Romaniello and P. L. de Boeij,
J. Chem. Phys. {\bf 122}, 164303 (2005).

\bibitem{marini.2001}
A. Marini, G. Onida, and R. Del Sole,
Phys. Rev. Lett. {\bf 88}, 016403 (2001).

\bibitem{marini.2002}
A. Marini, R. Del Sole, and G. Onida,
Phys. Rev. B {\bf 66}, 115101 (2002).

\bibitem{rohlfing.2010}
Z. Yi, Y. Ma, M. Rohlfing, V. M. Silkin, and E. V. Chulkov, 
Phys. Rev. B {\bf 81}, 125125 (2010).

\bibitem{gatti.2007}
M. Gatti, F. Bruneval, V. Olevano and L. Reining,
Phys. Rev. Lett, \textbf{99}, 266402 (2007).

\bibitem{bruneval.2006}
F. Bruneval, N. Vast, and L. Reining,
Phys. Rev. B {\bf 74}, 045102 (2006).

\bibitem{vanschilfgaarde.2006}
M. van Schilfgaarde, T. Kotani, and S. Faleev,
Phys. Rev. Lett. {\bf 96}, 226402 (2006).

\bibitem{kotani.2007}
T. Kotani, M. van Schilfgaarde, and S. V. Faleev,
Phys. Rev. B {\bf 76}, 165106 (2007);
T. Kotani, M. van Schilfgaarde, S. V. Faleev, and A. Chantis,
J. Phys.: Condens. Matter {\bf 19}, 365236 (2007).

\bibitem{heyd.2003}
J. Heyd, G. E. Scuseria and M. Ernzerhof,
J. Chem. Phys. {\bf 118}, 8207 (2003);
J. Chem. Phys. {\bf 124}, 219906 (2006).


\bibitem{heyd.2004}
J. Heyd and G. E. Scuseria, J. Chem. Phys. {\bf 120}, 7274 (2004);
J. Heyd and G. E. Scuseria, J. Chem. Phys. {\bf 121}, 1187 (2004).

\bibitem{heyd.2005}
J. Heyd, J. E. Peralta, G. E. Scuseria, R. L. Martin,
J. Chem. Phys. {\bf 123}, 174101 (2005).

\bibitem{peralta.2006}
J. E. Peralta, J. Heyd, G. E. Scuseria, and R. L. Martin,
Phys. Rev. B {\bf 74}, 073101 (2006).


\bibitem{itoh.1965}
T. Itoh, Rev. Mod. Phys. \textbf{37}, 159 (1965).

\bibitem{olevano.2010}
V. Olevano and M. Ladisa,
arXiv:11002.2117.

\bibitem{sakuma.2011}
R. Sakuma, C. Friedrich, T. Miyake, S. Blugel, and F. Aryasetiawan,
Phys. Rev. B, {\bf 84}, 085144 (2011).

\bibitem{supplemental}
See Supplemental Material at [URL will be inserted by publisher] for
convergence issues on the parameters involved in the $GW$ calculations.

\bibitem{hedin.1965}
L. Hedin,
Phys. Rev. {\bf 139}, A796 (1965);
L. Hedin and S. Lundqvist,
Solid State Phys. {\bf 23}, 1 (1969).

\bibitem{hybertsen.1985}
M. S. Hybertsen and S. G. Louie,
Phys. Rev. Lett. {\bf 55}, 1418 (1985).

\bibitem{hybertsen.1986}
M. S. Hybertsen and S. G. Louie,
Phys. Rev. B {\bf 34}, 5390 (1986).

\bibitem{godby.1986}
R. W. Godby, M. Schl\"uter, and L. J. Sham,
Phys. Rev. Lett. {\bf 56}, 2415 (1986);

\bibitem{godby.1988}
R. W. Godby, M. Schl\"uter, and L. J. Sham,
Phys. Rev. B {\bf 37}, 10159 (1988).

\bibitem{pulci.2005}
O. Pulci, M. Marsili, E. Luppi, C. Hogan, V. Garbuio, F. Sottile, R. Magri and R. Del Sole,
Phys. Stat. Sol. (b) {\bf 242}, 2737 (2005).

\bibitem{godby.1989}
R. W. Godby and R. J. Needs,
Phys. Rev. Lett. {\bf 62}, 1169 (1989).


\bibitem{aryasetiawan.2000}
F. Aryasetiawan, {\it Advances in Condensed Matter Science},
edited by I. V. Anisimov ~Gordon and Breach, New York, 2000.


\bibitem{kotani.2002}
T. Kotani and M. van Schilfgaarde,
Sol. Stat. Comm., {\bf 121}, 461 (2002).

\bibitem{lebegue.2003}
S. Lebegue, B. Arnaud, M. Alouani, and P. E. Bloechl,
Phys. Rev. B, {\bf 67}, 155208 (2003).

\bibitem{aryasetiawan.2008}
F. Aryasetiawan and S. Biermann,
Phys. Rev. Lett. {\bf 100}, 116402 (2008).

\bibitem{aryasetiawan.2009}
F. Aryasetiawan and S. Biermann,
J. Phys.: Condens. Matter {\bf 21}, 064232 (2009).

\bibitem{ashcroft.1966}
N. W. Ashcroft and N. D. Mermin,
{\it Solid State Physics}
(Holt, Rinehart and Winston, New York, USA, 1966).

\bibitem{note:lattice}
For the calculations with the hybrid functional, the lattice constant is taken to be the optimal DFT-PBE value of 7.90 Bohr.

\bibitem{abinit}
X. Gonze, B. Amadon, P.-M. Anglade, J.-M. Beuken, F. Bottin,
P. Boulanger, F. Bruneval, D. Caliste, R. Caracas, M. C\^ot\'e, T.
Deutsch, L. Genovese, P. Ghosez, M. Giantomassi, S. Goedecker,
D. Hamann, P. Hermet, F. Jollet, G. Jomard, S. Leroux, M.Mancini,
S. Mazevet, M. Oliveira, G. Onida, Y. Pouillon, T. Rangel,
G.-M. Rignanese, D. Sangalli, R. Shaltaf, M. Torrent, M. Verstraete,
G. Zerah and J. Zwanziger,
Comput. Phys. Commun.
{\bf 180}, 2582 (2009); http://www.abinit.org

\bibitem{vasp}
G. Kresse, J. Hafner, Phys. Rev. B {\bf 47}, RC558 (1993);
G. Kresse, J. Furthm\"uller, Phys. Rev. B {\bf 54}, 11169 (1996).

\bibitem{perdew.1996}
J. P. Perdew, K.Burke, M.Ernzerhof,
Phys. Rev. Lett. {\bf 77}, 3865 (1996).

\bibitem{teter.1993}
M. Teter, Phys. Rev. B {\bf 48}, 5031 (1993).

\bibitem{grinberg.2000}
I. Grinberg, N. J. Ramer, and A. M. Rappe,
Phys. Rev. B {\bf 62}, 2311 (2000) and
references therein.

\bibitem{note:pseudo}
For the fully relativistic DFT calculations, we use the Hartwigsen-Goedecker-Hutter~\cite{hartwigsen.1998} without semicore-states.

\bibitem{monkhorst.1976}
H. J. Monkhorst and J. D. Pack,
Phys. Rev. B {\bf 13}, 5188 (1976).

\bibitem{botti.2002}
S. Botti, N. Vast, L. Reining, V. Olevano and L. C. Andreani,
Phys. Rev. Lett. {\bf 89}, 216803 (2002).


\bibitem{shaltaf.2008}
R. Shaltaf, G.-M. Rignanese, X. Gonze, F. Giustino, and
A. Pasquarello,
Phys. Rev. Lett. {\bf 100}, 186401 (2008).

\bibitem{stankovski.2011}
M. Stankovski, G. Antonius, D. Waroquiers, A. Miglio, H. Dixit, K. Sankaran, M. Giantomassi, X. Gonze, M. C\^{o}t\'{e}, and G.-M. Rignanese, Phys. Rev. B {\bf 84}, 241201(R)(2011).

\bibitem{wannier90}
A. A. Mostofi, J. R. Yates, Y. Lee, I. Souza, D. Vanderbilt, and
N. Marzari,
Comput. Phys. Commun. {\bf 178}, 685
(2008); http://www.wannier.org

\bibitem{yates.2007}
J. R. Yates, X. Wang, D. Vanderbilt, and I. Souza,
Phys. Rev. B {\bf 75}, 195121 (2007).

\bibitem{hamann.2009}
D. R. Hamann and D. Vanderbilt,
Phys. Rev. B {\bf 79}, 045109 (2009).

\bibitem{rohlfing.1995}
M. Rohlfing, P. Kr\"uger, and J. Pollmann,
Phys. Rev. Lett. {\bf 75}, 3489 (1995).


\bibitem{olevano.2001}
V. Olevano and L. Reining,
Phys. Rev. Lett. \textbf{86}, 5962 (2001).

\bibitem{note:mixing}
It was found that the mixing of the occupied bands and the five first unoccupied bands resulting from the QS$GW$ procedure is negligibly modifying the initial DFT 5$d$-bands,
like in $G_0W_0$.

\bibitem{koelling.1977}
D. D. Koelling and B. N. Harmon,
J. Phys. C {\bf 10}, 3107 (1977).

\bibitem{takeda.1978}
T. Takeda,
Zeitschrift f\"ur
Phys. B Cond. Matt. and Quanta {\bf 32}, 43 (1978).

\bibitem{gollisch.1978}
H. Gollisch and L. Fritsche,
Phys. Status Solidi {\bf 86}, 145 (1978).

\bibitem{delsole.2003}
R. Del Sole, G. Adragna, V. Olevano and L. Reining,
Phys. Rev. B {\bf 67}, 045207 (2003).

\bibitem{stankovski-nonloc}
M. Stankowski, Local and non-local vertex corrections beyond the
GW approximation., Ph.D. thesis, University
of York (2008) ;
M. Stankovski, A. Morris, B. Robinson, R. Godby,
K. Delaney, P. Rinke, U. von Barth, C. Almbladh,
http://meetings.aps.org/link/BAPS.2007.MAR.U21.2
(2007).


\bibitem{rangel.2011}
T. Rangel,
{\it Many-body perturbation theory and Maximally-localized Wannier functions: a combined tool for first-principles electronic structure and quantum transport calculations},
Ph.D thesis, Universit\'e Catholique de Louvain (2011).



\bibitem{cazzaniga}
M. Cazzaniga, N. Manini, L. G. Molinari, and G. Onida,
Phys. Rev. B {\bf 77}, 035117 (2008).

\bibitem{mills.1980}
K. A. Mills, R. F. Davis, S. D. Kevan, G. Thornton,
and D. A. Shirley,
Phys. Rev. B {\bf 22}, 581 (1980).



\bibitem{baalmann.1983}
A. Baalmann, M. Neumann, H. Neddermeyer, W. Radlik, and
W. Braun,
Ann. Israel Phys. Soc {\bf 6}, 351 (1983).

\bibitem{jaklevic.1982}
R. C. Jaklevic and L. C. Davis,
Phys. Rev. B {\bf 26}, 5391 (1982).

\bibitem{szczepanek.1975}
P. Szczepanek and R. Glosser,
Solid State Commun. {\bf 15}, 4145 (1975).

\bibitem{jaklevic.1975}
R. C. Jaklevic and J. Lambe,
Phys. Rev. B {\bf 12}, 4146 (1975).

\bibitem{chen.1976}
A. Chen and B. Segall,
Solid State Commun. {\bf 18}, 149 (1976).

\bibitem{marel.1984}
D. van der Marel, G. A. Sawatsky, R. Zeller, F. U. Hillebrecht,
and J. C. Fuggle, Solid State Commun, {\bf 50}, 47 (1984).


\bibitem{jain.2011}
M. Jain, J. R. Chelikowsky and S. G. Louie, 
Phys. Rev. Lett. {\bf 107}, 216806 (2011).

\bibitem{paier.2006}
J. Paier, M. Marsman, K. Hummer, {\it et al.}
J. Chem. Phys. {\bf 124}, 154709 (2006).

\bibitem{marsman.2008}
M. Marsman, J. Paier, A. Stroppa, {\it et al.}
J. Phys. Condens. Matter {\bf 20}, 064201
(2008).


\bibitem{hartwigsen.1998}
C. Hartwigsen, S. Goedecker, and J. Hutter, Phys. Rev. B {\bf 58}, 3641 (1998).



\end{thebibliography}

\end{document}